# INCLUSIVE JETS AT THE TEVATRON[a]

SALLY SEIDEL

*The New Mexico Center for Particle Physics*
*The University of New Mexico*
*Albuquerque, NM 87131 USA*

*for the CDF and D0 Collaborations*

Results are presented for the inclusive jet cross section versus jet $E_T$ in $p - \overline{p}$ collisions at $\sqrt{s} = 1.8$ TeV as measured by the CDF and D0 detectors at Fermilab's Tevatron collider. The data are compared to next-to-leading-order QCD predictions using different input parton distribution functions. The ratio of inclusive jet cross sections at $\sqrt{s} = 0.63$ TeV and $\sqrt{s} = 1.8$ TeV, versus jet $x_T$, is also presented and compared to QCD predictions.

## 1 Introduction

Jet distributions at colliders are interesting to study for several reasons, in particular because they can signal the existence of new phenomena, test QCD predictions, and validate parton distribution functions. The complementary Tevatron detectors D0[1] and CDF[2] have studied the inclusive jet cross section at center-of-mass energies 1800 and 630 GeV.

The aspects of the two experiments that are especially important for the jet analyses are well described elsewhere[3]. For the results presented here, the radius of the reconstruction cone used[4,5] in both analyses is $R \equiv \sqrt{(\Delta\eta)^2 + (\Delta\phi)^2} = 0.7$, where $\eta$ is track pseudo-rapidity, $\phi$ is measured from the Tevatron plane, $\Delta\eta = \eta_2 - \eta_1$, $\Delta\phi = \phi_2 - \phi_1$, and the subscripts 1 and 2 correspond to the axis of the cone and the particle track, respectively.

## 2 The Inclusive Jet Cross Section Versus $E_T$ at $\sqrt{s} = 1800$ GeV

For jet transverse energies achievable at the Tevatron, the inclusive jet cross section probes distances down to $10^{-17}$ cm. For massless jets and $2\pi$ acceptance in $\phi$, this cross section, $Ed^3\sigma/dp^3$, can be written as the product:

$$E\frac{d^3\sigma}{dp^3} = \frac{1}{2\pi E_T}\frac{d^2\sigma}{dE_T d\eta}.$$

---

[a] Published in the Proceedings of the XXIX International Symposium on Multiparticle Dynamics (ISMD99), Providence, RI, 1999.

The second factor can be written in terms of the natural experimental variables:

$$\frac{d^2\sigma}{dE_T d\eta} = \frac{N}{\Delta E_T \Delta \eta L},$$

where $N$ is the number of jets observed, $\Delta E_T$ is the transverse energy bin size (5–80 GeV), $\Delta \eta$ is the pseudorapidity bin size (1.2), and $L$ is the luminosity.

CDF and D0 begin their analyses with similar data quality requirements. Both define the $z$-axis as the direction of the proton beam and place a cut on the absolute value of the $z$-coordinate of the primary vertex in the event in order to maintain the projective geometry of the calorimeter towers. CDF requires $|z_{\text{vertex}}| < 60$ cm; D0 requires $|z_{\text{vertex}}| < 50$ cm. To restrict the study to events whose energy is fully contained in the central barrel calorimeter, CDF (D0) requires that jets have pseudorapidity ($\eta_{\text{detector}}$) relative to the detector-based coordinate system such that $0.1 (0.0) \leq |\eta_{\text{detector}}| \leq 0.7 (0.5)$. To reject background due to accelerator loss, CDF further requires explicitly that the total energy in the event be less than 1800 GeV. Both experiments place cuts on the missing transverse energy ($\not{E}_T$) in the event in order to reject cosmic rays and mis-vertexed events. CDF requires that $\not{E}_T/\sqrt{\sum_{\text{all}} E_T} < 6$, while D0 requires that $\not{E}_T < 30$ GeV or $0.3 E_T^{\text{leading jet}}$, whichever is larger. Both experiments also place cuts on the ratio of energies detected by the electromagnetic and hadronic calorimeters and on jet shapes, in order to suppress background from noise.

The two experiments next correct for pre-scaling of triggers, detection efficiencies (these are typically in the range 94–100%), and "smearing,"[6,7] the last of which concerns the combined effect upon the data of energy mismeasurement and detector resolution. No correction is made for jet energy deposited outside the cone by the fragmentation process, as this is included in the next-to-leading order (NLO) calculation to which the data are ultimately compared.

Figure 1 shows the CDF measurement[8] of the inclusive jet cross section for collisions at $\sqrt{s} = 1.8$ TeV. This figure includes data measured in Runs 1a and 1b. The data are compared to the prediction by the NLO calculation by Ellis, Kunszt, and Soper (EKS)[9], in which the CTEQ4M[10] parton distribution function (PDF) has been used and the renormalization and fragmentation scales have both been set equal to $E_T/2$.

Figure 2 shows the quantity (DATA − THEORY)/THEORY for the same data for cases in which the PDF is CTEQ4M, CTEQ4HJ[10], and MRST[11]. Application to this analysis of the CTEQ4A and MRST PDF families has also been examined.

The effects on the cross section of a $1\sigma$ change in each of the CDF systematic uncertainties are shown in Figure 3. These uncertainties, which are fully correlated from bin to bin but are completely uncorrelated with each other, include the calorimeter's response to charged hadrons and showering particles, the stability of the energy scale, the details of the jet fragmentation model used in the simulation, the energy associated with the underlying event in the reconstruction cone, the modelling of the jet energy resolution function required for unsmearing, and the normalization. The excess at high values of $E_T$ that is present in the Run 1b data is consistent with what was previously observed in the Run 1a data. The analysis of the Run 1a excess has been described previously [12]. Quantitative comparison of the CDF Run 1b data with theory is now underway.

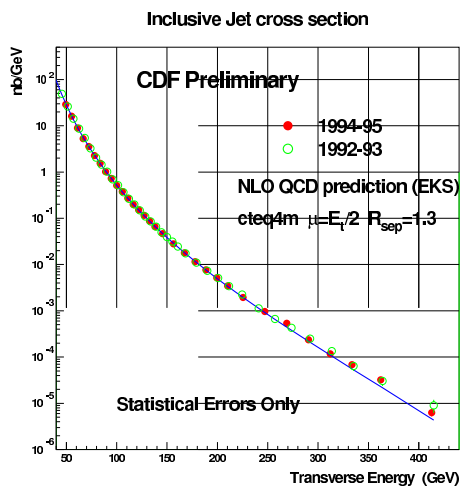

Figure 1: The preliminary measurement by CDF of the inclusive jet cross section for $\sqrt{s} = 1.8$ TeV, compared to the NLO EKS prediction with input parton distribution function CTEQ4M.

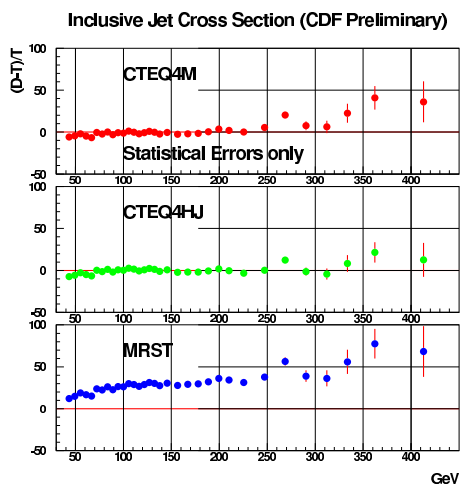

Figure 2: The percentage difference between the inclusive jet cross section as measured by CDF at $\sqrt{s} = 1.8$ TeV ("DATA"), and the EKS NLO prediction ("THEORY"), for a variety of input parton distribution functions.

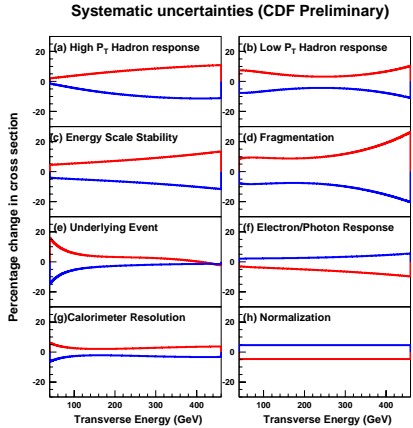

Figure 3: The percentage change in the Run 1b inclusive jet cross section, as measured by CDF, in response to a ±1 standard deviation change in each of the systematic uncertainties.

Figure 4 shows the D0 measurement [13] of the inclusive jet cross section. The data are compared with the NLO calculation JETRAD [14] with PDF CTEQ3M [15] and both scales set to one-half the maximum transverse energy associated with a jet in the event.

Figure 5 shows the quantity (DATA − THEORY)/THEORY for the same data for cases in which the PDF is CTEQ3M, CTEQ4M, and MRST.

The systematic uncertainties associated with the D0 measurement are displayed in Figure 6 and concern the calorimeter energy scale, the jet selection procedure, the uncertainties on trigger prescale values (denoted as the relative luminosity), the choice of jet energy resolution function used for unsmearing, and the luminosity. These uncertainties are all fully or partially correlated.

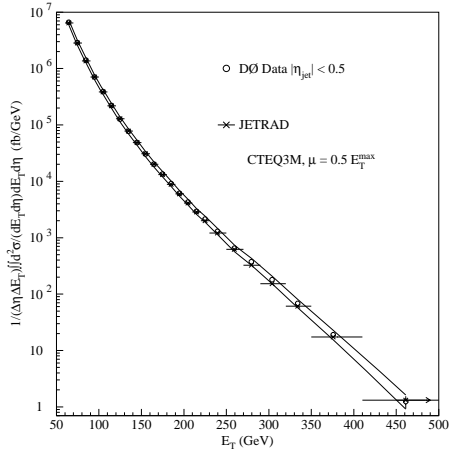

Figure 4: The measurement by D0 of the inclusive jet cross section for $\sqrt{s} = 1.8$ TeV, compared to the NLO prediction JETRAD with input parton distribution function CTEQ3M. The error bars indicate the statistical error, and the band represents the systematic.

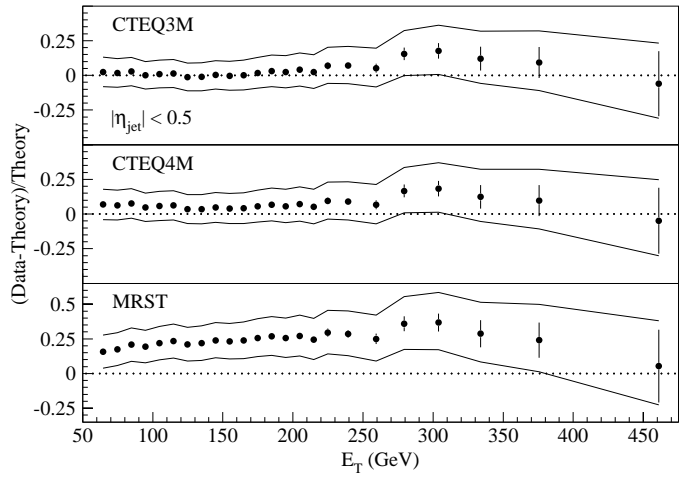

Figure 5: The percentage difference between the inclusive jet cross section as measured by D0 at $\sqrt{s} = 1.8$ TeV ("DATA"), and the JETRAD NLO prediction ("THEORY"), for a variety of input parton distribution functions.

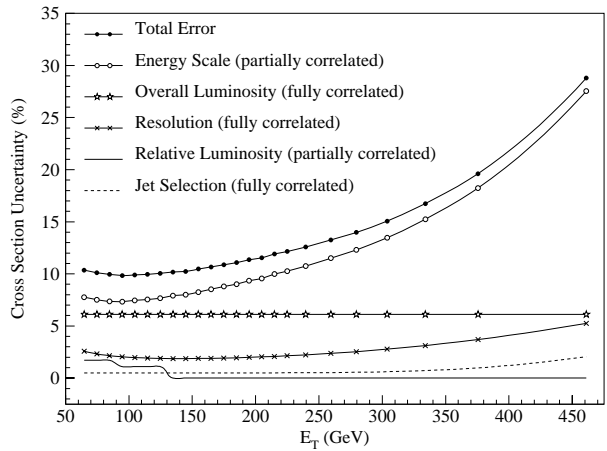

Figure 6: Contributions to the D0 inclusive jet cross section uncertainty.

The D0 Collaboration has conducted a comparison between the D0 data and theory. They define $\chi^2 \equiv \sum_{i,j}(D_i - T_i)(C^{-1})_{ij}(D_j - T_j)$, where $i$ is bin number, $D$ is the number of jets observed in the data, $T$ is the number of jets predicted by the theory, and $C$ is a covariance matrix which was constructed by analyzing the correlation of uncertainties between each pair of $E_T$ values. (Bin-to-bin correlations for representative $E_T$ bins are about 40% and positive.) There are 24 degrees of freedom (dof). Comparison of the data to the JETRAD calculation for 5 PDF's yields $\chi^2$/dof values that correspond to probabilities of agreement in the range 47–90% for $|\eta| \leq 0.5$, and 24–72% for $0.1 \leq |\eta| \leq 0.7$. Comparison of the D0 data to the EKS calculation using CTEQ3M, $R_{\mathrm{sep}} = 1.3R$, and scales $\mu = cE_T^{\mathrm{max}}$ or $cE_T^{\mathrm{jet}}$, for $c = 0.25$, $0.5$, and $1.0$, yield probabilities greater than or equal to 57% for all cases.

There is excellent agreement between the nominal CDF and D0 cross section values for $E_T \leq 350$ GeV. To quantify the level of agreement over the full $E_T$ range, D0 carried out a $\chi^2$ comparison between the D0 data and the nominal curve describing the central values of the CDF Run 1b data. The result was a $\chi^2$/dof $= 41.5/24$. As this comparison involved the nominal values of the CDF data, uncertainties on the CDF central values were not included. To approximate a comparison of the two data sets that includes information about the uncertainties on both, one can find the value of the CDF curve at each of the D0 $E_T$ points, multiply the D0 statistical errors by $\sqrt{2}$, and remove the 2.7% relative normalization difference, between the two experiments, as its origin is understood; this yields a $\chi^2$ of 35.1. One can then add systematic error information by expanding the covariance matrix to include both D0 and CDF uncertainties—this yields a $\chi^2$ of 13.1, corresponding to a probability of agreement of 96%.

## 3  The Inclusive Jet Cross Section Versus $E_T$ at $\sqrt{s} = 630$ GeV

One can multiply both sides of the inclusive jet cross section formula by $E_T^4$ to obtain the dimensionless cross section, $\sigma_d$, which is defined as

$$\sigma_d \equiv E_T^4 (E \frac{d^3\sigma}{dp^3}).$$

One can also define the scaled transverse energy of a jet, $x_T \equiv 2E_T/\sqrt{s}$. While the Naive Parton Model predicts that $\sigma_d$ is independent of $\sqrt{s}$ when plotted versus $x_T$ (the hypothesis of scaling), QCD predicts scaling violation due to the energy scale dependence of the probability for gluon radiation from a primary parton in the collision. A comparison of $\sigma_d$ measured at two different center-of-mass energies by the same experiment suppresses many theoretical and experimental uncertainties.

CDF and D0 collected 576 nb$^{-1}$ and 537 nb$^{-1}$ of data, respectively, at $\sqrt{s} = 630$ GeV. (Results from a data set of 8.6 nb$^{-1}$ collected at $\sqrt{s} = 546$ GeV were published [7] by CDF previously.) The data taken at $\sqrt{s} = 630$ and 1800 GeV were analyzed by the same method, the only difference in the analyses being the treatment of the correction for the energy of the underlying event, as this correction is known to increase with $\sqrt{s}$.

Figure 7 shows the results of the studies [16,17] with both sets of data normalized to theoretical calculations; in the case of D0, JETRAD is used, while in the case of CDF, EKS. Both calculations take the PDF MRSA′ [18].

The CDF systematic errors are shown separately in Figure 8. The CDF and D0 measurements agree with each other above about 80 GeV. While final conclusions must await further studies of the systematic errors, preliminary results indicate that the data may diverge in the lowest few $E_T$ bins. While the measurements are consistent with each other above 80 GeV, the theoretical calculation is somewhat higher for $E_T < 80$ GeV. Additional studies are needed for energy scale determination at low $E_T$ before definitive conclusions can be drawn.

## 4  The Ratio of the Dimensionless Cross Sections Versus $x_T$

Figure 9 shows the ratio of dimensionless cross sections versus jet $x_T$ as measured [17] by D0, compared to the JETRAD prediction for 7 combinations of

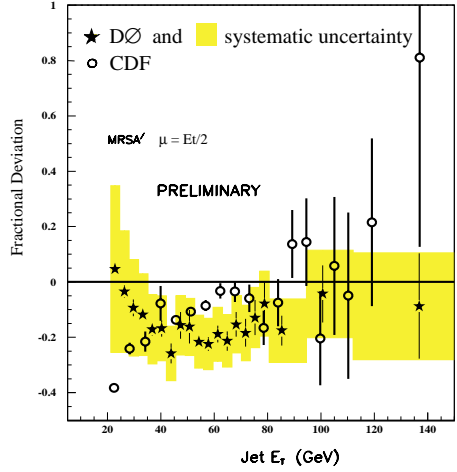

Figure 7: Preliminary D0 and CDF cross sections measured at $\sqrt{s} = 630$ GeV and compared to NLO QCD predictions. The shaded region indicates the D0 systematic errors.

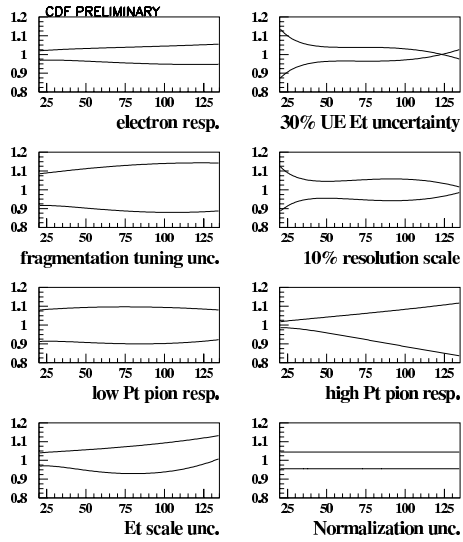

Figure 8: The percentage change in the inclusive jet cross section, as measured by CDF at $\sqrt{s} = 630$ GeV, in response to a $\pm 1$ standard deviation change in each of the systematic uncertainties.

PDF and scale. In each case the same value is used for the renormalization and the factorization scale at both values of $\sqrt{s}$. In all of these cases, the probability that the data and the theory are consistent lies in the range 0.01–7.2%.

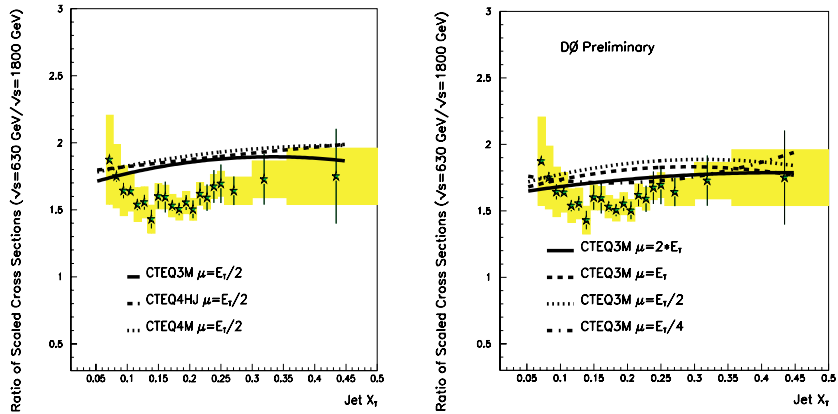

Figure 9: The preliminary measurement by D0 of the ratio of dimensionless cross sections taken at $\sqrt{s} = 630$ and 1800 GeV, compared to the NLO JETRAD prediction, for various combinations of parton distribution functions and scales.

Figure 10 shows the fractional error on the ratio per bin in jet $x_T$ and the bin-to-bin correlation of those errors.

D0 has also considered the case in which the scales are $\sqrt{s}$-dependent. They find, for example, that the combined choice of $\mu = 2E_T$ at $\sqrt{s} = 630$ GeV and $\mu = E_T/2$ at $\sqrt{s} = 1800$ GeV produces a prediction that has 95% probability of consistency with the data. D0 interprets this possible preference of the data for two different scales as an indicator that the next-to-next-to-leading-order terms in the matrix element for this process, when calculated, may not be negligible.

Figure 11 shows the CDF measurement [16] of the ratio of dimensionless cross sections versus jet $x_T$. Both the 630 GeV and the 546 GeV data are presented and are consistent. The measurements are compared to the EKS prediction for four choices of PDF and scale. In all four cases the same scale is used at both values of $\sqrt{s}$. The systematic errors on the CDF measurement are shown in Figure 12.

The ratio measurements by CDF and D0 are consistent for values of jet $x_T$ greater than 0.1. The discrepancy between the two data sets below that point may be traced to the measurement of $\sigma_d$ at $\sqrt{s} = 630$ GeV and was also apparent in the $\sqrt{s} = 546$ GeV data. There is, in addition, a slight overall

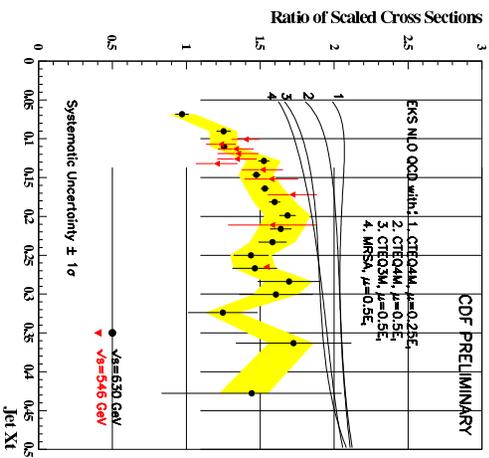
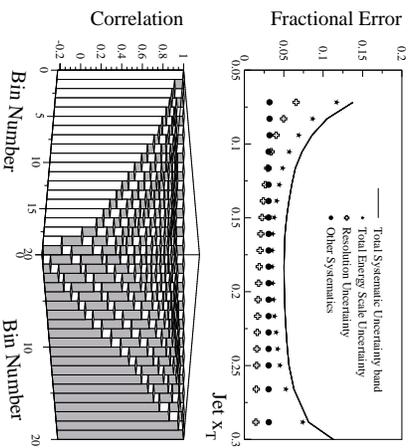

Figure 10: Top: the fractional error per $x_T$ bin on the D0 measurement of the inclusive jet cross section at $\sqrt{s} = 630$ GeV; bottom: the bin-to-bin correlation on the errors shown in the upper plot.

Figure 11: The preliminary measurement by CDF of the ratio of dimensionless cross sections taken at $\sqrt{s} = 630$ and 1800 GeV, compared to the EKS NLO prediction, for various combinations of parton distribution functions and scales.

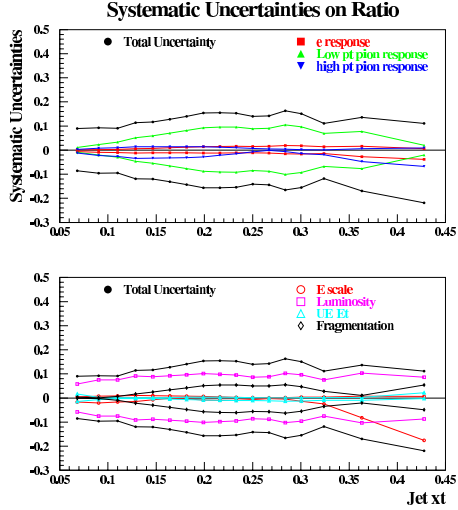

Figure 12: The percentage change in the ratio of dimensionless cross sections, as measured by CDF, in response to a ±1 standard deviation change in each of the systematic uncertainties.

normalization difference of about 20% between the theoretical predictions and the measurements.

### Acknowledgements

The author wishes to thank members of the CDF and D0 Collaborations, particularly Brenna Flaugher, Gerald Blazey, John Krane, Alexander Akopian, Robert Hirosky, Joey Huston, Andrew Brandt, Anwar Bhatti, and Greg Snow, for their advice on the presentation of these results.